\input amstex
\documentstyle{amsppt}
%
% Authors:  I. Yu. Cherdantzev, R. A. Sharipov
% Title:    Solitons on a finite-gap background
%           in Bullough-Dodd-Jiber-Shabat model
% Comments: AmSTeX, 7 pages, amsppt style, 1 figure
% E-mail:   R_Sharipov@ic.bashedu.ru
%           r-sharipov@mail.ru
% Web-page: http://www.geocities.com/r-sharipov
%
\leftline{International Journal of Modem Physics A,
Vol\.~5, No\.~15(1990)
3021-3027}
\leftline{\copyright\ World Scientific Publishing Company}
\vskip 2cm
\nopagenumbers
\TagsOnRight
\def\Res{\operatornamewithlimits{Res}}
\def\sign{\operatorname{sign}}
\def\const{\operatorname{const}}
\def\Real{\operatorname{Re}}
\def\Div{\operatorname{Div}}
\def\Prym{\operatorname{Prym}}
\def\negskp{\hskip -2pt}
\pagewidth{360pt}
\pageheight{606pt}
\rightheadtext{SOLITONS ON A FINITE-GAP BACKGROUND \dots} 
\topmatter
\title SOLITONS ON A FINITE-GAP BACKGROUND IN 
BULLOUGH-DODD-JIBER-SHABAT MODEL
\endtitle
\author
I.~YU.~CHERDANTZEV and R.~A.~SHARIPOV 
\endauthor
\abstract
The determinant formula for $N$-soliton solutions of the
Bullough-Dodd-Jiber-Shabat equation on a finite-gap
background is obtained. Nonsingularity conditions for
them and their asymptotics are investigated.
\endabstract
\address\vtop{\hsize=200pt\noindent Department of Mathematics,
\newline Bashkir State University,\newline Frunze str\.~32,
\newline 450074 Ufa,\newline Bashkiria,\newline USSR
\vskip 10pt\noindent
E-mail: \vtop{\hsize=150pt\noindent
R\_\hskip 1pt Sharipov\@ic.bashedu.ru\newline
r-sharipov\@mail.ru}\vskip 10pt\noindent
URL: http:/\negskp/www.geocities.com/r-sharipov}
\endaddress
\endtopmatter
\loadbold
\document
\vskip -20pt\noindent
{\bf 1. Introduction}
\vskip 10pt
Found in the early works on an inverse scattering method \cite{1, 2}
the Bullough-Dodd-Jiber-Shabat equation\footnotemark
\footnotetext{S.~P.~Tsarev discovered that this equation was first
found by Tzitzeica in \cite{12}. Now it is called Tzitzeica equation.
See also \cite{13} for more details.} 
$$
\hskip -2em
u_{xt}=e^u-e^{-2u}
\tag1.1
$$
makes a good alternative to the well-known Sine-Gordon equation as
one more model of the two-dimensional integrable relativistic field
theory with the self-action. In \cite{3} the authors had suggested
the construction of the finite-gap solutions for this equation.
These are periodic and/or almost periodic function in $x,t$ associated
with special classes of double sheet ramified coverings of Riemann
surfaces and represented by explicit formulae in terms of Prym
theta-functions of such coverings. In this paper we construct Hirota
type (see \cite{4}) determinant formula for $N$-soliton solutions of
\thetag{1.1} and use it for the investigation of the space-time
asymptotics of them in space-time variables where \thetag{1.1} has
a form
$$
\hskip -2em
u_{tt}-u_{xx}=e^u-e^{-2u}.
\tag1.2
$$
It is known that Eqs\.~\thetag{1.1} and \thetag{1.2} have no
fast-decreasing soliton solutions being the same time real-valued and
nonsingular. The presence of finite-gap background, as shown below,
makes it possible for nonsingular solitons to decrease rapidly to that
background. This has a transparent explanation in connection with the
topology of appropriate Riemann surface.
\vskip 10pt\noindent
{\bf 2. Main Construction}
\vskip 10pt
   First we touch briefly on the construction of finite-gap solutions
of \thetag{1.1} being the background in the following considerations
(for more details, see \cite{3}). Equation \thetag{1.1} has "zero
curvature" representation $[\partial_x-L,\,\partial_t-A]=0$ found in
\cite{5} with the matrix Lax operators of the form
$$
\xalignat 2
&\hskip -2em
L=\Vmatrix -u_x & 0 & \lambda\\
1 & u_x & 0\\ 0 & 1 & 0\endVmatrix,
&&A=\Vmatrix 0 & e^{-2u} & 0\\
0 & 0 & e^u\\ \lambda^{-1}\,e^u & 0 & 0
\endVmatrix.
\tag2.1
\endxalignat
$$
According to the general scheme of finite-gap integration (see review
\cite{6}) one should find the Baker-Achiezer vector function
$\psi=e(x,t,P)$ solving the spectral equations
$$
\xalignat 2
&\hskip -2em
\psi_x=L\psi,
&&\psi_t=A\psi
\tag2.2
\endxalignat
$$
and depending on a point $P$ of some compact Riemann surface $\Gamma$.
As soon as matrices $A$ and $L$ are not of general form, the so-called
``reduction problem'' arises. In \cite{3} we managed to solve it in our
particular case in a way similar to that of \cite{7} and \cite{8} where
the reduction of the general Schrodinger operator in magnetic field to
the purely potential Schrodinger operator was found. Let $\Gamma$ be the
compact Riemann surface of the even genus $g$ with the meromorphic
function $\lambda(P)$ having the only pole $P_\infty$ of the third order
and the only zero $P_0$ of the third order. In addition, let us suppose
that $\Gamma$ admits a holomorphic involution $\sigma$ with the property
$\lambda(\sigma P)=-\lambda(P)$ and the antiholomorphic involution $\tau$
with property $\lambda(\tau P)=\overline{\lambda(P)}$. For this case, we can
choose the local parameters $k^{-1}(P)$ and $q^{-1}(P)$ in the
neighborhoods of marked points $P_\infty$ and $P_0$ defined by the
following conditions $k^3=\lambda$, $q^{-3}=\lambda$, $k(\tau P)=\overline{
k(P)}$, $q(\tau P)=\overline{q(P)}$. Baker-Achiezer functions satisfying
\thetag{2.1} then are defined by fixing their essential singularities at
$P_\infty$ and $P_0$
$$
\xalignat 2
&\hskip -2em
\Vmatrix\psi_1\\ \psi_2\\ \psi_3\endVmatrix
\sim\Vmatrix k^{-1}\\ k^{-2}\\ k^{-3}\endVmatrix\cdot e^{kx},
&&\Vmatrix\psi_1\\ \psi_2\\ \psi_3\endVmatrix
\sim\Vmatrix q\,e^{-u}\\ q^2\,e^u\\ q^3\endVmatrix\cdot e^{qt}
\tag2.3
\endxalignat
$$
and by fixing the divisor $D$ of their poles. Divisor $D$ of degree $g$
should satisfy the following limitations
$$
\xalignat 2
&\hskip -2em
D+\sigma D-P_0- P_\infty\sim C,
&&\tau D=D,
\tag2.4
\endxalignat
$$
where $C$ is a divisor of canonical class on $\Gamma$. The above
limitations imposed on the choice of $\Gamma$ and $D$ make sure
that the matrices $L$ and $A$ are of the form \thetag{2.1} and
enable us to find the explicit formula for appropriate finite-gap
solution $u=v(x,t)$ of Eq\.~\thetag{1.1} in terms of the Prym
theta-function $\mu(z)$ of the covering r $\Gamma\to\Gamma/\sigma$:
$$
\hskip -2em
e^{v(x,t)}=c(\Gamma)-2\,\partial_{xt}\ln\mu(Ux+Vt+z_D).
\tag2.5
$$
Vector $z_D$ is determined by the divisor $D$, constant $c(\Gamma)$
depends only on Riemann surface $\Gamma$. Vectors $U$ and $V$ are
determined by the normalized Abelian differentials of the second
kind $\Omega_\infty$ and $\Omega_0$, the main parts of them at
$P_\infty$ and $P_0$ are of the form $\Omega_\infty=dk+\ldots$
and $\Omega_0=dq+\ldots$ respectively.\par
    Let $\Gamma$ be the $M$-curve with respect to anti-involution
$\tau$ (see \cite{9}). In this case condition \thetag{2.4} is
compatible with the choice of $D$ such that each invariant cycle
of $\tau$ contains only one point of $D$ except the cycle containing
$P_\infty$ and $P_0$ where there is no point of divisor $D$. It is
the very choice which provides the real and nonsingular finite-gap
solutions of Eq\.~\thetag{1.1} in \thetag{2.5}.\par
     In order to construct the $N$-soliton solutions of \thetag{1.1}
let us fix an extra set of parameters consisting of
\roster
\rosteritemwd=5pt
\item"(a)" a set of numbers $\lambda_1,\,\ldots,\,\lambda_N$ such that
$\lambda_i^2\neq\lambda_j^2$ for $i\neq j$;
\item"(b)" a set of nonzero constants $C_1,\,\ldots,\,C_N$.
\endroster
For general $\lambda_i$, each equation $\lambda(P)=\lambda_i$, has
exactly three solutions. Let us denote two of them as $\Lambda_i$,
and $\sigma\Lambda_i^*$, and then define the new $N$-soliton
Baker-Achiezer function $\Psi(x,t,P)$ by the following requirements:
\roster
\rosteritemwd=5pt
\item"(A)" $\Psi(x,t,P)$ is analytic everywhere on $\Gamma$ except
at $P_\infty$, $P_0$ and at points of the divisor $D+\Lambda_1+\ldots
+\Lambda_n+\Lambda_1^*+\ldots+\Lambda_N^*$;
\item"(B)" it has essential singularities at $P_\infty$ and $P_0$ of
the same form \thetag{2.3} as a purely finite-gap Baker-Achiezer
function;
\item"(C)" divisor $\Cal D=D+\Lambda_1+\ldots+\Lambda_n+\Lambda_1^*
+\ldots+\Lambda_N^*$ is its divisor of poles and
$$
\hskip -2em
\aligned
&\Res_{P=\Lambda_j}\bigl(3\,\Psi_i(P)\,\lambda^2(P)\,\omega(P)\bigr)
=C_j\cdot\Psi_i(\sigma\Lambda_j^*),\\
&\Res_{P=\Lambda_j^*}\bigl(3\,\Psi_i(P)\,\lambda^2(P)\,\omega(P)\bigr)
=-C_j\cdot\Psi_i(\sigma\Lambda_j).
\endaligned
\tag2.6
$$
\endroster
Here $\omega(P)=\omega_{0,\infty}(P)$ is an Abelian differential of
the third kind with the divisor of zeros and poles $D+\sigma D
-P_0-P_\infty$ and with the residue $+1$ at $P_0$ and the opposite
residue at $P_\infty$. The above conditions
\therosteritem{A}-\therosteritem{C} fix up the function
Baker-Achiezer $\Psi(x,t,P)$ uniquely. From (2.6) we have
$$
\aligned
\sum^N_{j=1}\ \Res\Sb\Lambda_j,\Lambda_j^*,\\
\sigma\Lambda_j,\sigma\Lambda_j^*\endSb
\bigl(\Psi_1(P)\,\Psi_2(\sigma P)\,\lambda(P)\,
\omega(P)\bigr)=0,\\
\vspace{1ex}
\sum^N_{j=1}\ \Res\Sb\Lambda_j,\Lambda_j^*,\\
\sigma\Lambda_j,\sigma\Lambda_j^*\endSb
\bigl(\Psi_3(P)\,\Psi_3(\sigma P)\,\lambda^2(P)\,
\omega(P)\bigr)=0.
\endaligned
$$
This is the duality condition for $\Psi(P)$ and $\Psi(\sigma P)$
which provides the equations \thetag{2.1} for $N$-soliton
Baker-Achiezer functions.\par
    In order to find the explicit formula for $N$-soliton solution
of Eq\.~\thetag{1 1} we shall use the technique developed in
\cite{10} and \cite{11}. For each pair of vector functions $\psi$
and $\phi$ we define the pairing $\left<\psi\,|\,\phi\right>$ as
follows:
$$
\hskip -2em
\left<\psi\,|\,\phi\right>=-\psi_1\,\phi_2+\psi_2\,\phi_1
+\lambda(P)\,\psi_3\,\phi_3.
\tag2.7
$$
The following properties of the pairing \thetag{2.7} are checked
up immediately:
$$
\pagebreak
\hskip -2em
\gathered
\partial_x\left<e(P)\,|\,e(\sigma Q)\right>
=\bigl(\lambda(P)-\lambda(Q)\bigr)\,e_2(P)\,e_3(P),\\
\partial_t\left<e(P)\,|\,e(\sigma Q)\right>
=e^u\left(1-\frac{\lambda(P)}{\lambda(Q)}\right)
e_3(P)\,e_1(P).
\endgathered
\tag2.8
$$
Because of \thetag{2.8} the value of $\left<e(P)\,|\,e(\sigma P)\right>$
does not depend on $x$ and $t$. One can calculate this value explicitly:
$$
\hskip -2em
\left<e(P)\,|\,e(\sigma P)\right>=\frac{d\lambda}{3\,\lambda^2\,\omega}.
\tag2.9
$$
Consider the following function depending on $x$, $t$ and on the pair of
points $P$ and $Q$ on Riemann surface $\Gamma$:
$$
\hskip -2em
\Omega(x,t,P,Q)=\frac{\left<e(x,t,P)\,|\,e(x,t\sigma Q)\right>}
{\lambda(P)-\lambda(Q)}.
\tag2.10
$$
When $Q$ is fixed function $\Omega$ has the pole divisor $D+Q$ and
essential singularities at $P_0$ and $P_\infty$ of the form
$$
\hskip -2em
\aligned
&\Omega(P,Q)\sim -\frac{q^2\,e^{qt}\,e_1(\sigma Q)}{\lambda(Q)}
\text{\ \ as \ }q\to\infty,\\
\vspace{2ex}
&\Omega(P,Q)\sim -k^{-4}\,e^{kx}\,e_2(\sigma Q)
\text{\ \ as \ }k\to\infty.
\endaligned
\tag2.11
$$
Moreover from \thetag{2.10} we derive
$$
\hskip -2em
\Res_{P=Q}\bigl(3\,\Omega(x,t,P,Q)\,
\lambda^2(P)\,\omega(P)\bigr)=1.
\tag2.12
$$
Properties of $\Omega(x,t,P,Q)$ just mentioned give us the
opportunity to use it for finding one of the components of
$N$-soliton Baker-Achiezer function $\Psi_3(x,t,P)$. We
$\Psi_3(x,t,P)$ construct via the following ansatz
$$
\hskip -2em
\gathered
\Psi_3(x,t,P)=e_3(x,t,P)+\sum^N_{j=1}\Omega(x,t,P,\Lambda_j)
\cdot\alpha_j(x,t)\,+\\
+\sum^N_{j=1}\Omega(x,t,P,\Lambda_j^*)\cdot\alpha_j^*(x,t)
\endgathered
\tag2.13
$$
with the parameters $\alpha_j$, $\alpha_j^*$ yet undefined. We shall
define them with the use of \thetag{2.6} by applying technique from
\cite{11}. $N$-soliton solution $u(x,t)$ of Eq\.~\thetag{1.1} then
is found on a base of equivalences
$$
\xalignat 2
&\partial_t e_3(P)\sim e^v\,k^{-4}\,e^{kx},
&&\partial_t\Psi_3(P)\sim e^u\,k^{-4}\,e^{kx}
\endxalignat
$$
as $P\to P_\infty$. Taking \thetag{2.11} into account, one gets
$$
\hskip -2em
e^u=e^v-\partial\left(\,\shave{\sum^N_{j=1}}e_2(\sigma\Lambda_j)
\cdot\alpha_j+\shave{\sum^N_{j=1}}e_2(\sigma\Lambda_j^*)
\cdot\alpha_j^*\right).
\tag2.14
$$
In general, to find $a_j$ and $a_j^*$ from \thetag{2.6} we should
solve some system of linear differential equations, the special
form of sums on the right part of \thetag{2.14} however makes it
possible to eliminate this step and leads us to the formula
$$
\hskip -2em
e^u=e^v-\partial_{xt}\ln\det(1-\Omega\,C).
\tag2.15
$$
Here $C=diag(C_1,\ldots,C_N,-C_1,\ldots,-C_N)$ is a diagonal matrix
matrix built of constants $C_j$ from \therosteritem{b} (see above).
Matrix $\Omega$ is defined by $\Omega(x,t,P,\Lambda_j$) and
$\Omega(x,t,P,\Lambda_j^*)$ with $P=\sigma\Omega_j$ and $P=\sigma
\Lambda_j^*$. Therefore its components could be evaluated explicitly
via Riemann theta functions. This matrix is composed of blocks
$$
\hskip -2em
\Omega=
\vcenter{\offinterlineskip
\def\vr{\vrule height 14pt depth 8pt}
\settabs\+\hskip 0.5cm&\hskip 3.7cm&\hskip 0.5cm&\hskip 3.7cm&\cr
\hrule
\+\vr\strut &$A_{ij}=\Omega(x,t,\sigma\Lambda_i,\Lambda_j)$
 &\vr\strut &$B_{ij}=\Omega(x,t,\sigma\Lambda_i,\Lambda_j^*)$
 &\vr\strut\cr
\hrule
\+\vr\strut &$C_{ij}=\Omega(x,t,\sigma\Lambda_i^*,\Lambda_j)$
 &\vr\strut &$D_{ij}=\Omega(x,t,\sigma\Lambda_i^*,\Lambda_j^*$
 &\vr\strut\cr
\hrule}
\tag2.16
$$
Formula \thetag{2.15} with the matrix \thetag{2.16} is a direct
analog of Hirota's determinant formula from \cite{4}.
\vskip 10pt\noindent
{\bf 3. Problem of Nonsingularity}
\vskip 10pt
    In order to get real and nonsingular $N$-soliton solutions
in \thetag{2.15} one should set more restrictions when choosing
parameters in \therosteritem{a} and \therosteritem{b}. Let the
Riemann \vadjust{\vskip 12pt\hbox to 0pt{\kern 62pt
\vbox{\special{em:graph pic1.gif}}\hss}\vskip 250pt
\centerline{\vbox{\hsize=200pt\noindent Fig\.~1. The quarter of
the Riemann surface $\Gamma$ unfolded onto the plane, $\tau$ acts
as a reflection downward, $\sigma$ acts as a rotation around the
point $P_\infty$.}}\vskip 12pt}surface
$\Gamma$ and divisor $D$ be chosen so that they correspond to
the real and nonsingular finite-gap solution $v(x,t)$ of
Eq\.~\thetag{1.1}. In this case surface $\Gamma$ can be cut and
then unwrapped onto a plane as shown on Fig\.~1. Points where
$\lambda(P)$ is purely real are shown with solid lines while
purely imaginary values of $\lambda(P)$ are shown with dashed
lines. A set of $\lambda(P)$ values on canonic $a$-cycles
(i\.\,e\. invariant cycles of $\tau$ except cycle $a_0$ containing
$P_0$ and $P_\infty$) consists of $g$ intervals. When $\lambda_j$
are in that interval the canonic choice $\Lambda_j$, $\Lambda_j^*$
pairs preserving $\lambda(\Lambda_j)=\lambda(\sigma\Lambda_j^*)>0$
is possible. Each pair $\Lambda_j$, $\Lambda_j^*$ together with the
point $P_i$, of divisor $D$ define the orientation $i_j$ on a cycle
$a_i$, in the direction from $\Lambda_j$ to $\sigma\Lambda_j$ through
$P_i$. Thus we can prescribe the sign to Abelian differential
$\omega(P)$ with respect to orientation $i_j$ just introduced. Let us
choose constants $C_j$, from \therosteritem{b} being real with the
signs defined by
$$
\hskip -2em
\sign(C_j)=\sign(\omega(P)|i_j)
\,\hbox{\vrule height 8pt depth 8pt width 0.5pt}_{\,P=\Lambda_j}.
\tag3.1
$$
It is easy to check that this rule gives an opposite sign on an
opposite cycle $\sigma a_i$ being in accordance with \thetag{2.6}.
\proclaim{Lemma 3.1} When sign rule \thetag{3.1} holds, the number
of zeros of the $N$-soliton Baker-Achiezer function $\Psi(P)$ on each
cycle $a_i$ is equal to the number of its poles on that cycle.
\endproclaim
Since $\Psi(P)$ is real on $a_i$, \thetag{2.6} and \thetag{3.1} and
simple geometrical considerations yield us the statement of lemma.
This means that all zeros of $\Psi(P)$ are on $a$-cycles, hence they
are separated from $P_0$ and $P_\infty$, i\.\,e\. they never coincide
nor come close to those points. The last fact in turn leads us to
the following theorem.
\proclaim{Theorem 3.1} The above canonical choice of $\Lambda_j$,
$\Lambda_j^*$ and the sign rule \thetag{3.1} for real constants
$C_j$ are enough to provide nonsingular $N$-soliton solution of
\thetag{1.1} in \thetag{2.15}.
\endproclaim
    Such facts are well-known in the theory of finite-gap
integration (see \cite{9}). In the case of fast-decreasing
solitons $\Gamma$ has a topology of sphere with the only
cycle $a_0$ containing $P_0$ and $P_\infty$, and as a result
we can find no place on $\Gamma$ to set $\Lambda_j$ and
$\Lambda_j^*$ as required by the theorem above.\par
    Formula \thetag{2.15} together with the following estimates
for purely finite-gap Baker-Achiezer function $e(x,t,P)$
$$
\hskip -2em
|e(x,t,P)|<\const\cdot\exp\bigl(\kappa_\infty(P)\,x
+\kappa_0(P)\,t\bigr)
\tag3.2
$$
via the Abelian integrals
$$
\xalignat 2
&\kappa_\infty(P)=\Real\int^P\Omega_\infty,
&&\kappa_0(P)=\Real\int^P\Omega_0
\endxalignat
$$
give us the opportunity to find asymptotics for $N$-soliton
solutions of Eq\.~\thetag{1.1} as $x,t\to\pm\infty$. Because
of \thetag{3.2} we may divide entire rows and columns of
matrix $\Omega$ in \thetag{2.15} by exponential factors
of growth
$$
\exp\bigl(2(\kappa_\infty(\sigma\Lambda_j^*)-\kappa_\infty(
\sigma\Lambda_j))\,x+2(\kappa_0(\sigma\Lambda_j^*)-\kappa_0(
\sigma\Lambda_j))\,t\bigr).
$$
Vanishing of $(\kappa_\infty(\sigma\Lambda_j^*)-\kappa_\infty(
\sigma\Lambda_j))\,x+2(\kappa_0(\sigma\Lambda_j^*)-\kappa_0(
\sigma\Lambda_j))\,t$ determines the free soliton trajectories.
Their velocities are
$$
\pagebreak
\hskip -2em
v_j=-\frac{\kappa_\infty(\sigma\Lambda_j^*)-\kappa_\infty(\sigma
\Lambda_j)}{\kappa_0(\sigma\Lambda_j^*)-\kappa_0(\sigma\Lambda_j)}
\tag3.3
$$
For Eq\.~\thetag{1.2} in space-time coordinates soliton velocities
$V_j$, are bound with $v_j$ as
$$
\hskip -2em
V_j=\frac{e^\varepsilon\,v_j+e^{-\varepsilon}}
{e^{-\varepsilon}-e^\varepsilon\,v_j}
\tag3.4
$$
Here $\varepsilon$ is a scalar parameter depending on the
choice of particular system of Lorentzian coordinates. The
matter that $V_j$ are less than light velocity $c=1$ here
is a consequence of \thetag{3.3} and \thetag{3.4} and the
following lemma.
\proclaim{Lemma 3.2} When points $\Lambda_j$, $\sigma\Lambda_j^*$
are placed as shown on Fig\.~1, the inequalities $\kappa_\infty(\sigma
\Lambda_j^*)-\kappa_\infty(\sigma\Lambda_j),0$ and $\kappa_0(\sigma
\Lambda_j^*)-\kappa_0(\sigma\Lambda_j)<0$ hold.
\endproclaim
    The observers being at rest in almost all Lorentzian systems
of coordinates will see the similar pictures of soliton interaction:
before and after the interaction solitons are free. They are separated
by finite-gap background potentials $v(x,t,z_j)$ with the common
spectrum $\Gamma$ but different phases $z_j$, peculiar to each
interval between solitons. The background phase shift obtained while
crossing the path of $j$-th soliton is $\triangle z_j=U(\Lambda_j
+\Lambda_j^*-\sigma\Lambda_j+\sigma\Lambda_j^*)$.
Here $\triangle z_j$ is a vector of Prym variety of the covering
$\Gamma\to\Gamma/\sigma$ and $U$ is a map from subset of $\Div(\Gamma)$
into $\Prym(\Gamma)$ built in paper \cite{3}.
\vskip 8pt
\rightline{Received 2 November 1989}


\vskip 10pt\noindent
{\bf References}
\vskip 10pt
\Refs\nofrills{}
\ref
\no 1\by R.~K.~Bullough and R.~K.~Dodd\jour Proc\. R.~Soc\. London
\vol 352A\yr 1977\page 481
\endref
\ref
\no 2\by A.~V.~Jiber and A.~B.~Shabat\jour Dokl\. Akad\. Nauk SSSR
\vol 247\yr 1979\page 1103
\endref
\ref
\no 3\by I.~Yu\.~Cherdantzev and R.~A.~Sharipov\jour Theor\. Math\.
Phys\.\vol 82\yr 1990\page 155
\endref
\ref
\no 4\by R.~Hirota\jour Phys\. Rev\. Lett\.\vol 27\yr 1971\page 1192
\endref
\ref
\no 5\by A.~V.~Mikhailov\jour Physica~3D\vol 1\yr 1981\page 73
\endref
\ref
\no 6\by B.~A.~Dubrovin, I.~M.~Krichever, and S.~P.~Novikov
\inbook Seria ``Fundamentalnie napravlenia''\vol 4\yr 1985
\page 179
\endref
\ref
\no 7\by A.~P.~Veselov and S.~P.~Novikov\jour Dokl\. Akad\.
Nauk SSSR\vol 279\yr 1984\page 20
\endref
\ref
\no8\by A.~P.~Veselov and S.~P.~Novikov\jour Dokl\. Akad\.
Nauk SSSR\vol 279\yr 1984\page 784
\endref
\ref
\no 9\by B.~A.~Dubrovin and S.~M.~Natanson\jour Func\. Anal\.
i Pril\.\vol 16\yr 1982\page 27
\endref
\ref
\no 10\by R.~A.~Sharipov\jour Dokl\. Akad\. Nauk SSSR
\vol 292\yr 1987\page 1356
\endref
\ref
\no 11\by R.~A.~Sharipov\jour Usp\. Mat\. Nauk\vol 41
\yr 1986\page 202
\endref
\vskip 2pt
\hrule width 50pt
\vskip 3pt
\ref\no 12\by G.~Tzitzeica\jour Comptes Rendu de l'Acad\'e\-mie
des Sciences\yr 1910\vol 150\page 955\moreref\page 1227
\endref
\ref\no 13\by A.~Yu.~Boldin, S.~S.~Safin, and R.~A.~Sharipov
\jour Journ\. of Math\. Phys\.\yr 1993\vol 34\issue 12\page 5801
\endref
\endRefs
\enddocument
\end